# Failure mechanisms of single-crystal silicon electrodes in lithium-ion batteries

Feifei Shi[1,2], Zhichao Song[1], Philip N. Ross[2], Gabor A. Somorjai[2,3], Robert O. Ritchie[1,2,4] & Kyriakos Komvopoulos[1]

Long-term durability is a major obstacle limiting the widespread use of lithium-ion batteries in heavy-duty applications and others demanding extended lifetime. As one of the root causes of the degradation of battery performance, the electrode failure mechanisms are still unknown. In this paper, we reveal the fundamental fracture mechanisms of single-crystal silicon electrodes over extended lithiation/delithiation cycles, using electrochemical testing, microstructure characterization, fracture mechanics and finite element analysis. Anisotropic lithium invasion causes crack initiation perpendicular to the electrode surface, followed by growth through the electrode thickness. The low fracture energy of the lithiated/unlithiated silicon interface provides a weak microstructural path for crack deflection, accounting for the crack patterns and delamination observed after repeated cycling. On the basis of this physical understanding, we demonstrate how electrolyte additives can heal electrode cracks and provide strategies to enhance the fracture resistance in future lithium-ion batteries from surface chemical, electrochemical and material science perspectives.

[1] Department of Mechanical Engineering, University of California, Berkeley, California 94720, USA. [2] Materials Sciences Division, Lawrence Berkeley National Laboratory, Berkeley, California 94720, USA. [3] Department of Chemistry, University of California, Berkeley, California 94720, USA. [4] Department of Materials Science and Engineering, University of California, Berkeley, California 94720, USA. Correspondence and requests for materials should be addressed to R.O.R. (email: roritchie@lbl.gov) or to K.K. (email: kyriakos@me.berkeley.edu).

Despite significant research devoted to the exploration of new types of batteries[1–3], lithium-ion batteries (LIBs) remain the most extensively used power source for various applications, such as portable electronics, electric vehicles and long-term energy storage. In common with traditional batteries, LIBs have two electrodes that reversibly host lithium-ion insertion and extraction[4]. Novel electrode materials, such as silicon, have been proposed as a promising upgrade for the current graphite carbon-based electrodes because of their improved gravimetric energy density[5]. However, the larger capacity density implies that the silicon electrode must host more insertion and extraction of lithium ions during lithiation/delithiation cycling. As a result, the silicon electrode experiences excessive volume expansion and contraction cyclically, which induces irreversible electrode deformation and fracture. Consequently, the mechanical degradation of the silicon electrode results in severe capacity and power fade, thereby greatly limiting the battery's long-term durability for critical applications, such as power systems of electric vehicles. Accordingly, here we seek to provide a fundamental understanding of the failure mechanisms of silicon electrodes in LIBs over extended cycles, to provide guidance for new and improved electrode design with minimal capacity decay.

It has been suggested that decreasing electrode material size to nanostructures in the form of nanoparticles, nanowires or nanotubes is a promising strategy for preventing electrode failure[5–7]. In addition, critical sizes of 'fracture-free' nanoparticle and nanowire electrodes have actually been proposed[8,9]. These delicate nanostructured silicon electrodes, however, are still not ready for commercialization due to several reasons. First, the coulombic efficiency of these nanostructures in the first cycle is naturally very poor because their high surface-to-volume ratio causes more lithium ions to become trapped in the solid electrolyte interphase (SEI). Second, the much higher manufacturing cost of nanostructured electrodes may offset the competitive advantages associated with their presumed improved capacity. Third, the size of silicon particles that commercial vendors can manufacture with reasonable control is still larger than the estimated 'fracture-free' size. Consequently, the salient mechanisms controlling the fracture and hence the lifetime of silicon electrodes of LIBs remain an unavoidable issue that must be understood. To this end, various theoretical models have been developed and experimental studies performed to elucidate the fracture mechanisms of both crystalline and amorphous silicon electrodes[10–13]. However, there have been few attempts to investigate the physical mechanisms underlying fracture behaviour of these electrodes at more realistic larger scales (that is, from micrometres to centimetres), especially during long-term cycling[14].

In this study, we combine chemical and electrochemical experiments with fracture mechanics and the finite element method (FEM) to investigate the electrochemical and mechanical response of single-crystal silicon electrodes subjected to long-term cycling. Single-crystal silicon is chosen because it provides an ideal model surface and bulk material; moreover, as standard electrochemistry measurements can be readily made, it is possible to track the development of a crack in the electrode and, most importantly, identify its trajectory over extended cycles. The simulated stress/strain contours and predicted progress of the crack paths are shown to be consistent with the experimental observations. Specifically, with increasing number of cycles, perpendicular cracks initiate at the electrode surface, propagate in the vertical (thickness) direction and eventually deflect along the lithiation boundary causing delamination. We believe that these results have significant implications to the understanding of the progressive failure of silicon electrodes and provide guidance for the development of design strategies that can mitigate the degradation and failure of silicon electrodes in LIBs.

## Results

**Voltammetry and evolution of electrode surface morphology**. Figure 1a–c shows cyclic voltammetry curves of a p-type boron-doped Si(100) electrode subjected to 30 cycles of voltage between 2.0 and 0.01 V at a scan rate of 0.1 mV $s^{-1}$. (All the electrochemical tests were performed in a custom-made reaction cell shown in Supplementary Fig. 1.) The onset of the reduction current is observed in the first cycle at $\sim$0.1 V, corresponding to the initial alloying of the crystalline silicon with lithium, while in the second cycle the lithiation reduction peak shifts to $\sim$0.3 V (Fig. 1a). The two anodic peaks at $\sim$0.3 and $\sim$0.5 V are associated with Li dissolution[15–17]. The magnitude of the oxidation and reduction peaks increases with the number of cycles (Fig. 1b), which can be attributed to the continuous increase of the amount of electrochemically active silicon in each cycle. The kinetics of lithiation in silicon is analogous to the model of $SiO_2$ layer formation on silicon[18]. During the initial cycles (Fig. 1a,b), the increase of active silicon is controlled by the lithiation reaction rate; thus the current density increases with the number of cycles (that is, lithiation time). Both the lithiation depth and the time needed for lithium to transport from the surface to the reaction front increase with cycling. After 30 cycles, the time for lithium transport approaches the cycle time and the cyclic voltammetry curve stabilizes (Fig. 1c). With continued cycling of the silicon electrode, the redox current area begins to decrease, indicating the loss of active silicon material.

Here we characterize the evolution of the surface morphology of the silicon electrode using top-view scanning electron microscope (SEM) images obtained after 3, 8 and 50 cycles (Fig. 1d–f) and magnified SEM images of the electrode surface obtained after 30 cycles (Fig. 1g–i). The dominant feature is the development of surface cracks in two orthogonal directions, resulting in isolated small squares that form a regular repetitive pattern on the electrode surface. These crack patterns further account for the formation of square cavities (Fig. 1f). Apparently, the damage to the electrode surface is not a rapid and catastrophic process, but occurs in a gradual and cumulative fashion. While damage of the electrode surface is insignificant during the first few cycles, shallow surface cracks initiate after a critical number of cycles and propagate with further cycling, eventually causing delamination at the electrode surface.

**Formation of orthogonal surface cracks**. To explain the origin of these orthogonal surface cracks, we need to consider that lithiation in single-crystal silicon is a strongly orientation-dependent process. Although the entry of lithium ions into the silicon electrode occurs perpendicular to the electrode surface (that is, the (100) direction), further lithium flux in the electrode bulk occurs in all directions. In particular, lithiation invasion preferentially occurs in the $\langle 110 \rangle$ direction and is significantly less in the $\langle 100 \rangle$ and $\langle 111 \rangle$ directions[19]. Figure 2a shows a top-view SEM image of $3 \times 3 \times 8\ \mu m^3$ micropillars fabricated by photolithography on a Si(100) substrate with exposed {110} lateral surfaces and round corners in the $\langle 100 \rangle$ direction. (Top-view and three-dimensional micropillar images are also shown in Supplementary Fig. 2.) With increasing lithiation, the straight edges of the micropillars preferentially expand in the $\langle 110 \rangle$ directions (Fig. 2b–d), while the round corners expand significantly less. This anisotropic expansion makes the 'rotation' of the micropillars in the $\langle 100 \rangle$ direction to vanish (Fig. 2d,e) and the neighbouring micropillars to merge (Fig. 2f). If a micropillar is presumed to represent a 'unit cell' of the solid

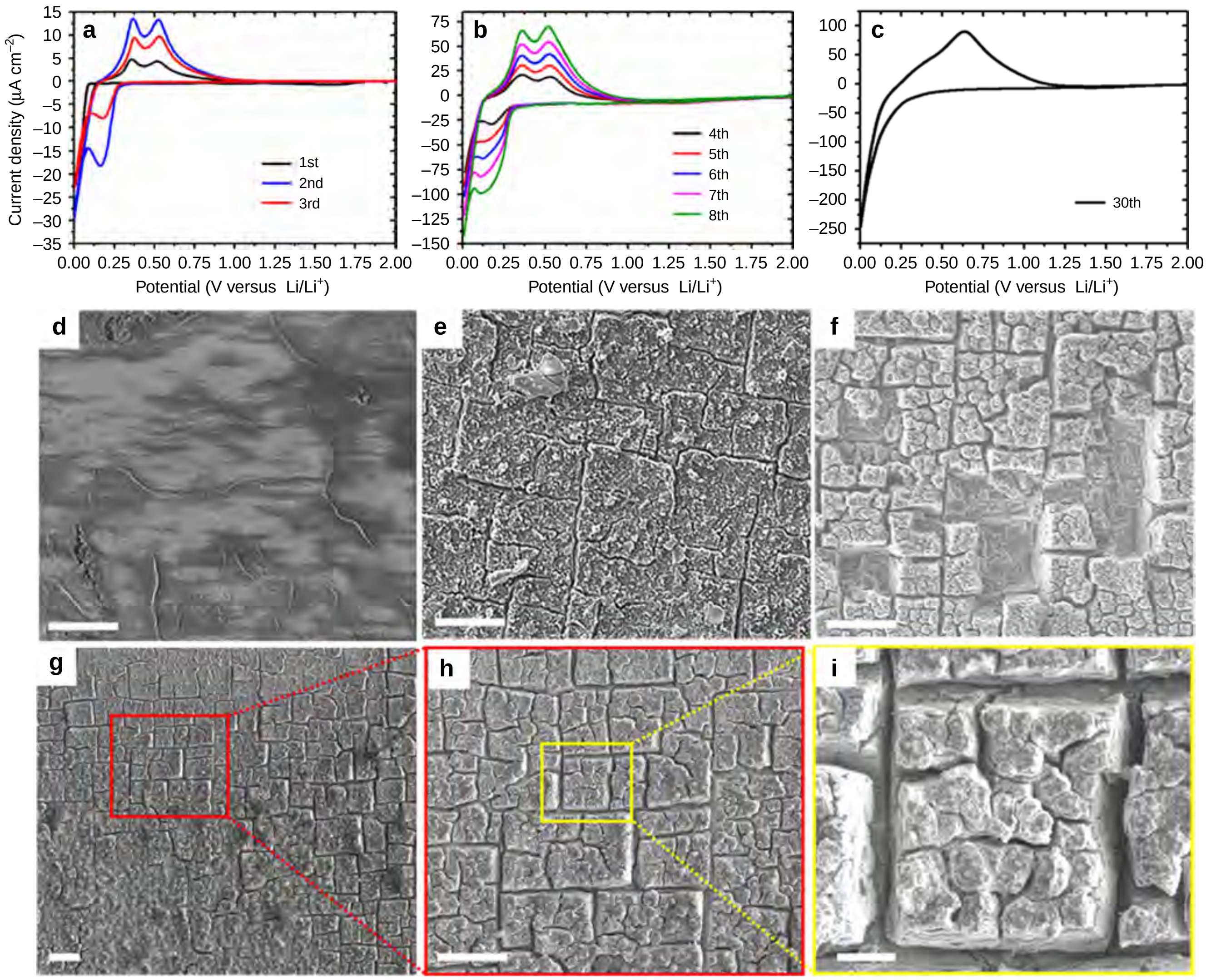

**Figure 1 | Electrode cyclic voltammetry response and surface morphology.** Current-voltage curves of a Si(100) electrode cycled between 2.0 and 0.01 V at a scan rate of 0.1 mV s$^{-1}$ for (**a**) 1-3, (**b**) 4-8 and (**c**) 30 successive lithiation/delithiation cycles. Top-view scanning electron microscope images of electrode surface morphology obtained after (**d**) 3, (**e**) 8, (**f**) 50 and (**g–i**) 30 cycles. Scale bar, 20 μm (**d–h**); 3 μm (**i**).

Si(100) electrode, the volumetric expansion during cycling will be confined by the surrounding material, resulting in the development of a compressive force at the micropillar edges. Because of the significantly higher lithiation rate in the <110> directions than the <100> directions, the 'unit cell' is subjected to a higher compressive force on all the {110} planes, which, in turn, induces a high shear stress and plasticity along the ±45°directions, that is, along the {100} planes.

To validate the above hypothesis, a planar octahedral model of the 'unit cell' of the silicon electrode was developed and analysed with the FEM (Fig. 2g). (More details about the 'unit cell' model and associated boundary conditions can be found in Supplementary Methods and related Supplementary Fig. 3.) Lithiation-induced volumetric expansion was simulated as a thermal expansion process. A moving temperature field was applied to model lithium invasion and migration into the silicon electrode. The unlithiated (crystalline) silicon (c-Si) was modelled as an isotropic elastic material of elastic modulus $E = 185$ GPa and Poisson's ratio $v = 0.22$, whereas the lithiated (amorphous) silicon (a-Si) was modelled as an elastic–plastic material with material properties depending on lithium-ion concentration (that is, $E = 50$–$102$ GPa, $v = 0.22$ and yield strength $Y = 0.5$–$3.0$ GPa)[13]. Because the pillar height is much larger than the in-plane pillar dimensions and the expansion in the (100) direction is less than that along the in-plane <110> directions, plane-strain conditions were assumed in all simulations. Although the axial elongation is not exactly zero, cross-sectional SEM images of the lithiated crystalline pillars revealed a negligible change in pillar height compared with in-plane expansion, providing experimental evidence of the validity of the plane-strain assumption. Figure 2g shows the deformed configuration of a partially lithiated micropillar, including contours of equivalent plastic strain. Due to the anisotropic lithiation rate (lithium invasion in the (110) direction is 6.4 times faster than that in the (100) direction[11]), the expansion is anisotropic, in agreement with our experimental observations (Fig. 2b,c) and those of others[10,18]. This anisotropic expansion leads, in turn, to the development of a high shear stress, which is responsible for crack initiation. Figure 2h shows the equivalent plastic strain in a 'unit cell' of the electrode after full lithiation. Because volumetric expansion of the 'unit cell' is fully constrained by the surrounding material, large plastic strains develop along the (100) edges, contributing to the formation of perpendicular stress bands along the <100> directions.

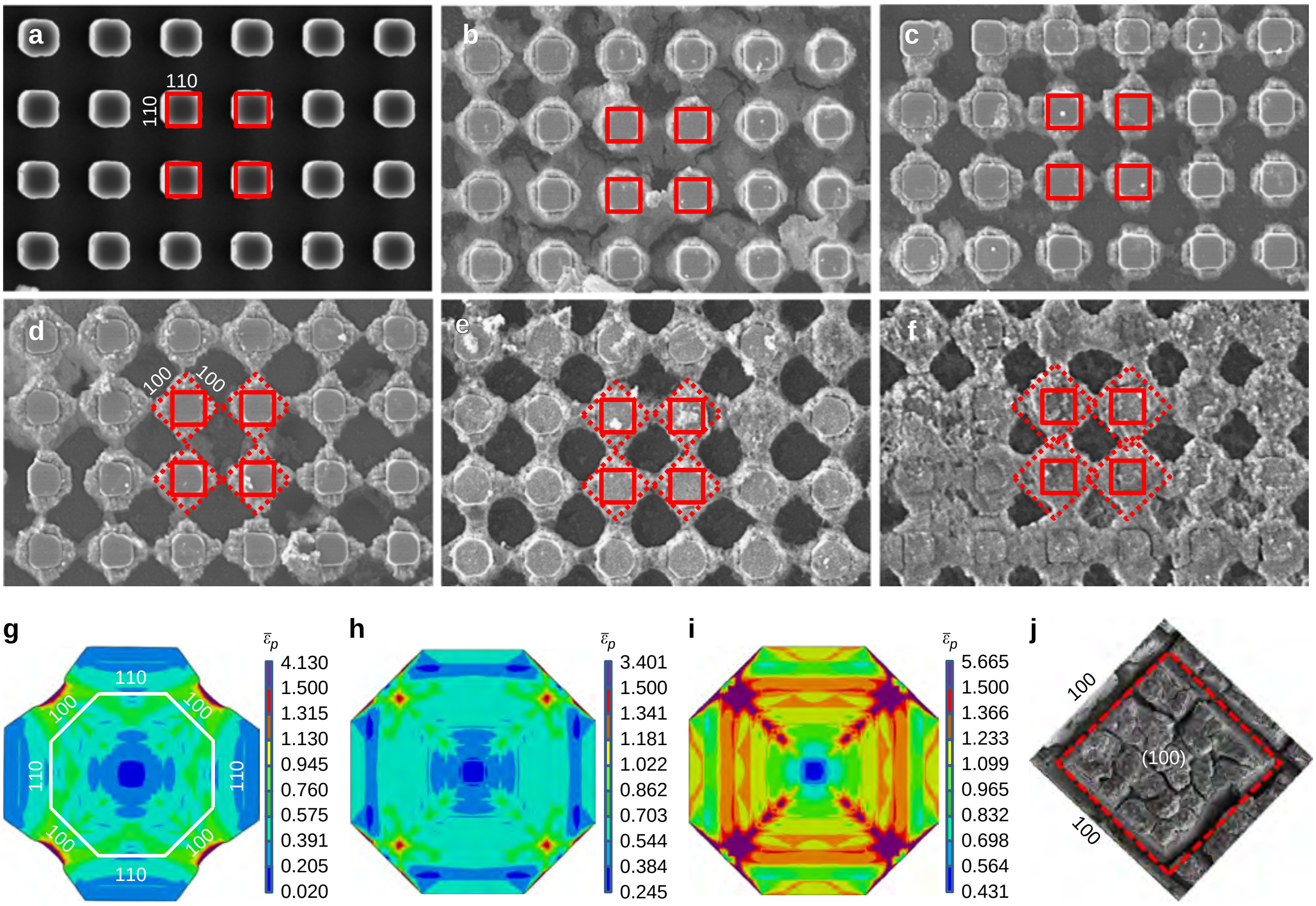

**Figure 2 | Anisotropic in-plane deformation and fracture of lithiated Si(100) micropillars and electrodes.** (**a**–**f**) Scanning electron microscope (SEM) images of square micropillars fabricated on a p-type Si(100) wafer obtained at various lithiation stages (the size of the red solid squares is $3 \times 3\,\mu m^2$). Finite element results of in-plane equivalent plastic strain $\bar{\varepsilon}_p$ distribution in (**g**) lithiated (unconstrained) micropillar and (**h**) lithiated and (**i**) delithiated (constrained) octahedral 'unit cell' of the silicon electrode. (**j**) SEM image showing the crack pattern on the surface of a lithiated silicon electrode revealing crack formation in the $<100>$ directions.

Figure 2i shows equivalent plastic strain contours after full delithiation, indicating that during delithiation the locations of large plastic strain continue to accumulate plasticity as a pair of perpendicular deformation bands. The field of the equivalent plastic strain in the electrode bulk can be assembled by packing the equivalent plastic strain contours in the 'unit cell' to construct the high-plastic-strain bands at larger scales, in agreement with the experimental observation of repeated perpendicular surface cracks at different length scales.

**Surface crack initiation, propagation and deflection**. After the formation of cracks perpendicular to the electrode surface (Fig. 1e,g–i), small squares of electrode material defined by these surface cracks begin to delaminate (Fig. 1f). To understand the origin of this process, it is necessary to consider how these surface cracks form and progressively propagate with cycling before causing electrode material to delaminate. This can be accomplished by analysing the experimental evidence of surface crack growth during the first 50 lithiation/delithiation cycles. Figure 3a–d (left) shows cross-sectional SEM images across a typical surface crack after 3, 8, 30 and 50 cycles, respectively. The crack initiates at the electrode surface (Fig. 3b), consistent with the fracture of single-crystal silicon nanopillars[19], propagates in the depth direction with further cycling (Fig. 3c) and, after reaching a depth of $\sim 5$–$8\,\mu m$, is deflected laterally (Fig. 3d). The reason for the formation of square delamination fragments of electrode material is the lateral deflection of neighbouring cracks after prolonged cycling (Fig. 1f).

Despite the clear evidence from cross-sectional SEM images of the observed crack trajectories, the question that remains is the physics and mechanics underlying such behaviour; specifically, why do cracks initiate at the electrode surface and why do such initially vertical cracks deflect laterally? To seek answers to these questions, we analysed the lithiation/delithiation-induced stress and strain fields in the cross-section of the silicon electrode over multiple cycles. A cohesive zone model (Supplementary Fig. 4; Supplementary Methods; Supplementary Note 1) was used to represent multiple interfaces, which were allowed to separate at a critical stress and thus simulate crack initiation and growth. These interfaces are characterized by the cohesive strength $\sigma_c$ and the interface work of adhesion $G_c$ (fracture energy). An approximately linear lithium concentration profile (Supplementary Fig. 5; Supplementary Note 2) was simulated by the gradual advancement of the lithiation boundary with increasing lithiation/delithiation cycles.

As shown in Fig. 3a (middle column), after three cycles both the lithiation depth $d$ ($\sim 1\,\mu m$) and plastic deformation are mainly confined at the electrode surface. With continued cycling, the lithiation depth increases with more plasticity accumulating below the surface owing to further lithium insertion. During the subsequent delithiation, the tensile residual stresses generated in the large plastic region induce face separation along the cohesive interface, resulting in the initiation and propagation of vertical cracks (Fig. 3b,c, right). When the crack tip approaches the lithiation boundary, the crack is abruptly deflected laterally and continues to propagate along the a-Si/c-Si interface (Fig. 3d, right), which is consistent with the experimental evidence (Fig. 3d, left). Such a marked deflection in crack trajectory can be interpreted in terms of the mutual competition between the direction of maximum mechanical driving force and the weakest microstructural path[20]. Specifically, from a mechanics perspective, cracks in nominally brittle (elastic) materials follow the path of maximum strain energy release rate. This is essentially consistent with a $K_{II}=0$ crack trajectory, where $K_{II}$ is the stress intensity factor for in-plane shear crack displacement, although this is mitigated in real materials by the nature of the microstructure encountered by the crack. In the current system pertaining to silicon electrodes of LIBs, although the elastic modulus of c-Si (185 GPa) is higher than that of a-Si (50–102 GPa), its fracture toughness $G_c \approx 9\,\mathrm{J\,m^{-2}}$ ($K_{IC} \approx 1.3\,\mathrm{MPa\,m^{1/2}}$) is an order of magnitude higher than that of the a-Si/c-Si interface in the (100) direction, whose estimated fracture toughness is $G_c \approx 1\,\mathrm{J\,m^{-2}}$ ($K_{IC} \approx 0.3\,\mathrm{MPa\,m^{1/2}}$)[21]; accordingly, the crack is deflected along the a-Si/c-Si interface, because this is both energetically more favourable and the weakest microstructural path.

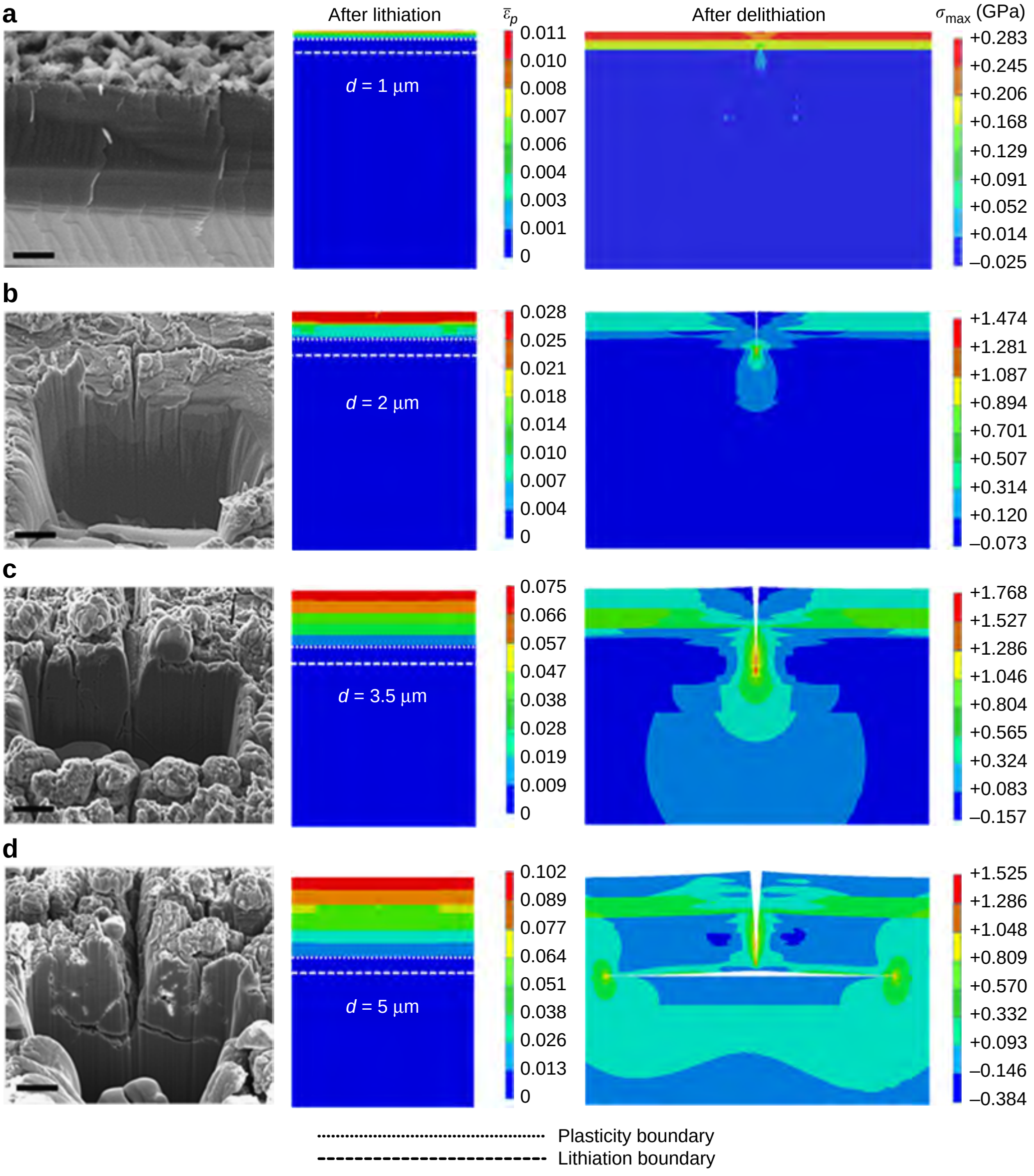

**Figure 3 | Experimental and numerical results of electrode fracture.** Cross-sectional focused ion beam (FIB)-scanning electron microscope (SEM) images (left column) and corresponding finite element method (FEM) results (middle and right columns) of a Si(100) electrode obtained after (**a**) 3, (**b**) 8, (**c**) 30 and (**d**) 50 lithiation/delithiation cycles. The FIB-SEM images show crack growth through the electrode thickness followed by crack deflection and propagation along the lithiation boundary (lithiated/unlithiated interface). The FEM results show the equivalent plastic strain $\bar{\varepsilon}_p$ after lithiation and the maximum principal stress $\sigma_{max}$ after delithiation, illustrating crack initiation, propagation through the electrode thickness and lateral deflection along the lithiation boundary for different lithiation/delithiation cycles. Scale bar, 0.5 µm (**a**); 3 µm (**b**–**d**).

**Fracture mechanics-based interpretation**. Such deliberations of the behaviour of cracks impinging on dissimilar (elastic) material interfaces can be explained in mechanics terms by the analysis of He and Hutchinson[22], which provides a quantitative criterion to predict whether a singular crack will penetrate through or deflect along a linear-elastic bimaterial interface. As depicted in Fig. 4, the crack path depends on two principal factors, namely, the elastic modulus mismatch across the interface, described by the Dundurs' parameter $\alpha = (E_1 - E_2)/(E_1 + E_2)$, where $E_1$ and $E_2$ are the respective elastic moduli of material 1 and 2, and the ratio of the toughness (critical strain energy release rate) of the interface to that of the material beyond the interface $G_{c,int}/G_{c,2}$. By substituting the material and interfacial properties of the present system (given in the preceding paragraph) into this analysis, we can calculate that the modulus mismatch $\alpha$ for the lithiated/unlithiated a-Si/c-Si interface is approximately equal to $-0.4$, with a toughness ratio $G_{c,int}/G_{c,2} \approx 0.11$ (red square point in Fig. 4). Accordingly, it is clear from Fig. 4 that, from a mechanical driving force perspective, the crack will definitely deflect along the a-Si/c-Si interface.

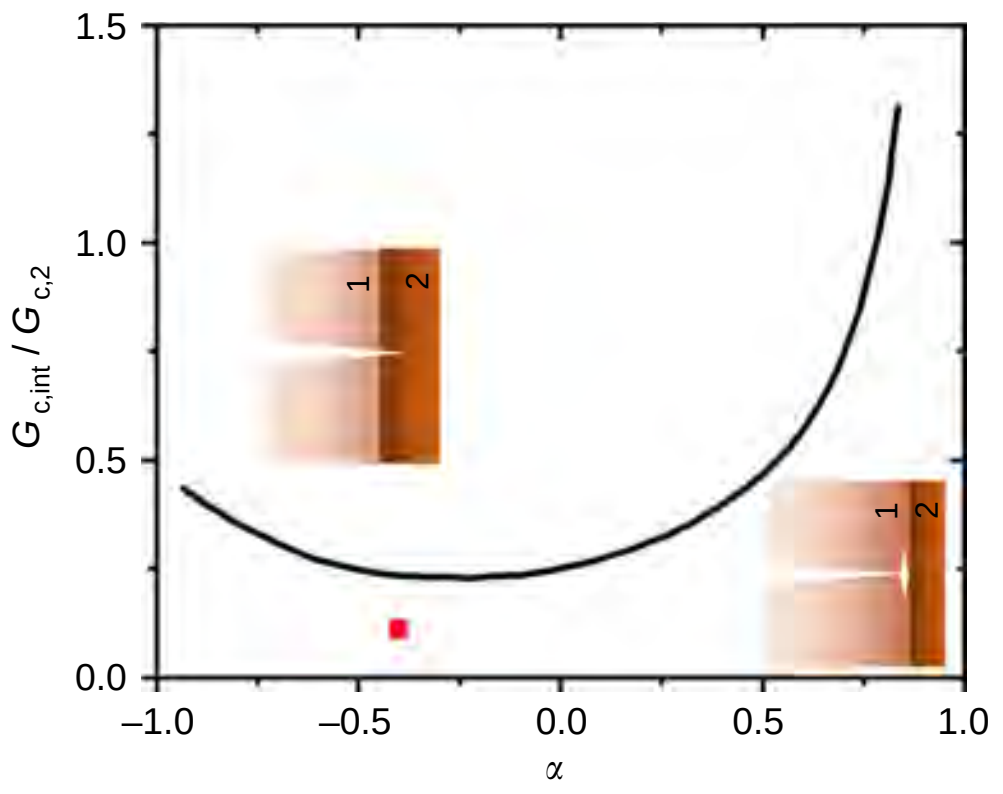

**Figure 4 | Crack deflection at a bimaterial interface.** Linear-elastic solution of interface to material beyond the interface toughness $G_{c,int}/G_{c,2}$ ratio versus Dundurs' parameter $\alpha = (E_1 - E_2)/(E_1 + E_2)$, where $E_1$ and $E_2$ are the elastic modulus of material 1 and 2, respectively[22]. Whether the crack penetrates through the interface or is deflected along the interface depends on the elastic modulus mismatch (represented by the Dundurs' parameter) and interface-to-material 2 (c-Si) toughness ratio. For the a-Si/c-Si interface, $\alpha \approx -0.4$ and $G_{c,int}/G_{c,2} \approx 0.11$ (red square point). The theory predicts that a crack growing in the a-Si will not propagate into the c-Si, but will be deflected along the a-Si/c-Si interface, in agreement with the experimental observations (Fig. 3, left column).

Inspired by the mechanistic understanding of the silicon electrode failure, we proceeded to investigate if electrolyte additive, which is typically used to modify the chemical properties of the SEI[23,24], can affect crack growth and, in turn, electrode degradation. A Fourier transform infrared (FTIR) spectroscopy investigation has shown that the reduction product of vinylene carbonate (VC) additive is poly-VC, whereas that of the fluoroethylene carbonate (FEC) additive is alkyl carbonate $ROCO_2Li$ salt[25–27]. Both of these reaction products can chemically passivate the electrode surface before the decomposition of the electrolyte solvent. The poly-VC aggregates fill the vertical surface cracks (Fig. 5a,b) and promote crack-face bridging (Fig. 5c). As a result, the fracture toughness is increased and further crack propagation is inhibited. The crack depth after 30 cycles is ~100 nm, which is significantly smaller than the 5–8 μm crack depths observed with additive-free electrolyte consisting of 1 M lithium hexafluorophosphate ($LiPF_6$) in ethylene carbonate (EC)/diethyl carbonate (DEC) (EC:DEC = 1:2 v/v). The $ROCO_2Li$ salt uniformly precipitates on the electrode surface (Fig. 5d,e), as evidenced by the roughening of the electrode surface. Although surface cracks are still visible, they do not perfectly align in perpendicular directions as for the additive-free electrolyte (Fig. 1g), because $ROCO_2Li$ continues to form on the exposed edges and faces (Fig. 5f). These $ROCO_2Li$ salts serve to fill and passivate the surface cracks, thereby significantly decelerating crack growth. We believe that these findings will stimulate research to formulate new electrolyte additives for enhancing the electrode fracture resistance.

## Discussion

In addition to the electrolyte additive strategy described above, the fracture mechanics model introduced in this study can also lead to other approaches for improving the electrode robustness and durability, including: (i) Surface modification (artificial SEI design). Recent FTIR spectroscopy studies[28–30] have shown that the main components of the SEI forming on silicon electrodes have high solubility in the electrolyte and very weak mechanical strength. A thin protective film of enhanced chemical stability, high fracture toughness and strong adhesion to the electrode surface can minimize the surface crack initiation by reducing the surface plasticity[7]. (ii) Electrode material selection. Despite its high capacity, the silicon electrode material is naturally brittle and prone to fracture. Emerging ductile electrode materials (for example, silicon-metal alloy) and structurally optimized electrode materials (for example, porous/hollow structure composite with tuned binder) may significantly reduce the risk of electrode fracture[6,31,32]. (iii) Charging profile optimization. Secondary ion mass spectrometry (SIMS) shows approximately linear decrease in lithium concentration through the electrode thickness (Supplementary Fig. 5). By tuning the charging profile, we can obtain the same capacity with a uniform lithium distribution through the electrode thickness. This would allow a lower surface lithium concentration, thereby reducing surface plasticity, which is a key factor for suppressing crack initiation and propagation.

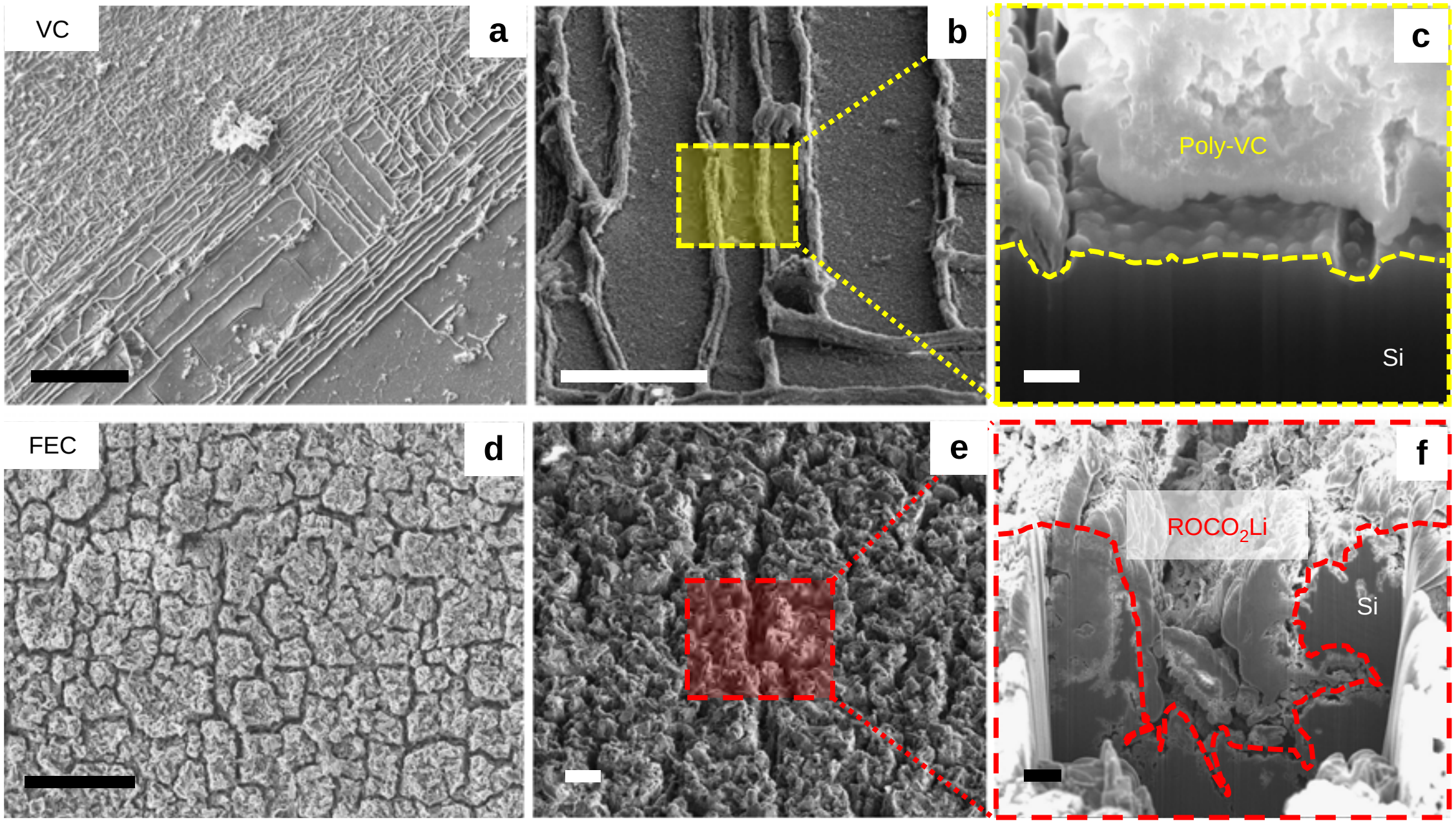

**Figure 5 | Electrode crack arrest by the reduction products of electrolyte additives.** Top-view and cross-sectional scanning electron microscope images of a Si(100) electrode cycled between 2.0 and 0.01 V for 30 cycles at a scan rate of 0.01 mV s$^{-1}$ in EC/DEC/lithium hexafluorophosphate ($LiPF_6$) electrolyte with (**a**–**c**) 5% vinylene carbonate (VC) additive and (**d**–**f**) 5% fluoroethylene carbonate (FEC) additive. The dashed rectangles shown in **b** and **e** indicate the focused-ion-beam milling area. The reduction products of the VC and FEI additives (poly-VC and $ROCO_2Li$, respectively) modify the mechanical properties of the solid electrolyte interphase (SEI) film and inhibit crack growth. Scale bar, 100 μm (**a**,**d**); 20 μm (**b**,**e**); 300 nm (**c**); 3 μm (**f**).

The present work provides a mechanistic explanation of the root cause of premature silicon electrode failure in LIBs following prolonged lithiation/delithiation cycling. On the basis of this insight, we demonstrate how electrolyte additives change the SEI, tune the mechanical properties of the cohesive layer and improve the electrode fracture resistance. In addition, the present failure analysis may also guide the optimization of other strategies, including surface modification, electrode material development and tuning of the charge profile to increase the material ductility and suppress the high surface stress. We consider the knowledge of the failure mechanisms of silicon electrodes gained from the present study as a foundation for future design improvements and the aforementioned strategies as main paths towards the development of more robust and durable electrodes for next-generation LIBs.

## Methods

**Electrode preparation.** The silicon electrodes used in this investigation were p-type boron-doped Si(100) wafers of 500 μm thickness and 0.001 Ω cm electrical resistivity (MTI Co.). The native oxide film was removed by first treating with diluted 5% hydrofluoric acid and then rinsing with ultrapure water of 18.2 MΩ cm electrical resistivity for 2 min. The $3 \times 3 \times 8\,\mu m^3$ micropillar electrodes used in this study (Supplementary Fig. 2) were fabricated by standard photolithography. Reactive ion etching was performed with the Bosch process.

**Electrochemical measurements.** All of the electrochemical experiments were performed in a single-compartment teflon cell (Supplementary Fig. 1). The electrolyte consisted of 1 M $LiPF_6$/EC/DEC (EC:DEC = 1:2 v/v) (Novolyte Technologies, BASF) with or without 5% VC or FEC additive. Dimethyl carbonate (DMC) (>99.9%, high-performance liquid chromatography grade, Sigma-Aldrich) was used as the rinsing solvent. The counter and reference electrodes were made of Li, whereas the working electrodes were cut from Si(100) wafers with or without micropillars. All potentials reported in this work are referred to the $Li/Li^+$ redox couple. All electrochemical tests were performed inside a glove box filled with Ar gas ($H_2O$ and $O_2$ contents <10 p.p.m.). A multi-channel potentiostat (Multistat 1480, Salartron Analytical) was used for potential/current control.

**Surface morphology characterization.** After electrochemical treatment, the silicon samples were first rinsed with DMC to remove the residual electrolyte and then transferred to an SEM (JSM-6700F, JEOL) for imaging. The electrode microstructure was observed with a focused ion beam (Nova 600i Dual Beam, FEI). The cycled electrodes were cross-sectioned with a $Ga^+$ ion beam and observed with the SEM.

**Secondary ion mass spectrometry.** SIMS measurements of the lithiated silicon electrodes were obtained by Evans Analytical Group. The Cameca IMS 4f magnetic sector of the SIMS instrument was operated at a pressure of $2.67 \times 10^{-7}$ Pa ($2 \times 10^{-9}$ torr). Elemental concentration depth profiles were obtained by analysing positively charged CsLi and CsSi secondary ions. The $Cs^+$ current was typically equal to 10 nA. The sputtering depth was determined from the depth of the sputtering craters measured with a profilometer. Data acquisition and post-processing analysis were performed with the SIMSview software of the SIMS instrument. The lithiated samples were sealed in a glove box and transferred to the SIMS spectrometer within 1 min.

**Finite element analysis.** The general purpose FEM code ABAQUS/Standard (version 6.14) was used to determine the in-plane stress and strain fields resulting from anisotropic lithiation/delithiation and to analyse crack initiation, propagation, deflection and delamination over multiple lithiation/delithiation cycles. The unlithiated silicon (c-Si) was modelled as an isotropic elastic material with $E = 185$ GPa and $\nu = 0.23$, whereas the lithiated silicon (a-Si) was modelled as an isotropic elastic-perfectly plastic material with elastic modulus and yield strength varying with the lithium-ion concentration.

**Data availability.** Experimental and numerical data supporting the findings of this investigation are available from the corresponding authors on request.

## Acknowledgements

This work was supported by the Assistant Secretary for Energy Efficiency and Renewable Energy, Office of Freedom CAR and Vehicle Technologies, US Department of Energy under contract no. DE-AC02 O5CH1123. K.K. also acknowledges the funding provided for this work by the UCB – KAUST Academic Excellence Alliance Program. The potentiostat instrumentation was purchased with funding from the Director, Office of Science, Office of Basic Energy Sciences, Materials Science and Engineering Division, US Department of Energy, which also provided support to R.O.R. We thank C. Shen for assistance in micropillar sample preparation.

## Authors contributions

F.S. designed and performed the experiments, Z.S. performed the FEM simulations, F.S., Z.S, P.N.R., G.A.S., R.O.R and K.K. analysed the results, F.S., Z.S, R.O.R and K.K. prepared and revised the manuscript, and K.K. supervised and directed the work of F.S. and Z.S. throughout this project.

## Additional information

**Supplementary Information** accompanies this paper at http://www.nature.com/naturecommunications

**Competing financial interests:** The authors declare no competing financial interests.

**Reprints and permission** information is available online at http://npg.nature.com/reprintsandpermissions/